\documentclass[twocolumn,prl,reprint,amsmath,amssymb,showpacs]{revtex4}
\usepackage{graphicx}

\begin{document}

\title{Cooperatively enhanced light transmission in cold atomic matter}

\author{Kasie Kemp, S.J. Roof, and M.D. Havey}

\affiliation{Old Dominion University, Department of Physics, Norfolk, Virginia 23529 \\
}

\author{I.M. Sokolov${}^{1,2}$ and D.V. Kupriyanov${}^{1}$}

\affiliation{$^{1}$Department of Theoretical Physics, State Polytechnic
University, 195251, St.-Petersburg, Russia \\ \small $^{2}$Institute for
Analytical Instrumentation, Russian Academy of Sciences, 198103, St.-Petersburg, Russia  }

\begin{abstract}
We report enhanced transmission in measurements of the spectral dependence of forward light scattering by a high-density and cold ensemble of $^{87}Rb$ atoms. This phenomenon, which is a result of dipole-dipole interaction induced cooperative light scattering in the atomic sample, implies a significant departure from the traditional density dependence of the transmitted light as embodied in the Beer-Lambert Law.  Absolute values of the density-dependent forward light scattering cross-section are extracted from the measurements.
\end{abstract}


\maketitle

The general optical properties of aggregated matter differ significantly from those of individual nuclear or atomic scatterers.  In the case of a dilute and optically thin gas of atoms the response of the system to applied monochromatic electromagnetic radiation is quite well understood.  However, upon increasing the atomic density, the quantum optics of such systems develops correlated characteristics \cite{Sokolov1,Sokolov2,Balik,Kaiser,Kupr}. This is especially true in cold atomic gases, where coherence shared between applied and scattered fields, and the atoms comprising the samples, can survive environmental decoherence for substantial periods of time. Among those phenomena mediated by spatial disorder in confined cold atomic samples, Anderson localization \cite{Anderson,SokolovSkipetrov} of light and random atomic lasing \cite{Kaiser1,Cao,Letokhov,Gerasimov} are areas of current investigation. Because of the global coherence that can be generated in cold atomic ensembles, an optically excited ensemble can be considered a single entity, and the system described by a collective of atomic superpositions. One possible representation of these is the super and subradiant states introduced by Dicke \cite{Dicke}. Although superradiant states have been observed and studied by many investigators, experimental research on the relatively fragile subradiant states has been much more limited.   Recent work, for instance, has shown that, through off-resonance optical excitation of an atomic system, a so-called timed-Dicke state may be created \cite{Scully1,Rohlsberger1}.   Such a collective excitation distributed through the sample as a whole, can demonstrate the possibility of one-photon superradiance. Further, the timed-Dicke state may be mixed through so-called Fano couplings into subradiant modes, which may then be experimentally observable.

To fundamentally describe such rich cooperative and collective phenomenology in cold gases requires quantitative understanding of their microscopic properties.  For optically excited homonuclear gases, the constituent particles interact on a wide range of length scales governed by angle-dependent $(kr)^{-n}$, (n = 1,2,3) terms in the pairwise radiative dipole-dipole interaction.  Here $r$ is the separation between a pair of atoms and $k = 2\pi / \lambda$ is the light wave vector.  The system as a whole is then representative of a near resonant many body system, and should be strongly modified from the response of its constituent parts.  A conceptual basis of collective quasimodes \cite{Balik}, is useful for understanding such conditions \cite{Sokolov1,Sokolov2,Balik}.  Comprehensive theoretical treatments of such systems \cite{Friedberg1,Manassah1} have predicted modification of the optical response, including a collective Lamb shift.   Such effects, including the customary Lorentz-Lorenz local field shift, have indeed been observed in warm vapors \cite{Maki,Keaveney,Kampen}.  On the other hand, a recent letter \cite{Pellegrino} has reported on the suppression of diffuse light scattering with increasing density in a dense and cold cloud of $^{87}Rb$ atoms.  These measurements also indicate a small frequency shift of the collective atomic resonance.  Theoretical research \cite{Juha,Juha2} suggests that such spectral features are emergent phenomena associated with mean field descriptions of the interaction of light and matter and appear only upon the inclusion of an inhomogeneous interaction.  Theoretical work \cite{Sokolov1,Sokolov2} on the transmission and dispersive properties of near resonance light incident on a cylindrical sample of stationary atoms has shown that there is a substantial enhancement of coherent light transmission over what would be expected according to the Beer-Lambert Law.  In this case, the normalized transmission T, measured as -lnT, is proportional to the optical depth b = $\rho \sigma L$; $\rho$ is the atomic density, $\sigma$ is the scattering cross section $\emph{for a free atom}$, and L is the sample length.  Such an effect was also predicted more recently by \cite{Kaiser}.

In this letter we report experimental investigation of forward scattering of near resonance radiation from a large and dense cloud of cold $^{87}Rb$ atoms. We investigate the dependence of the forward scattered intensity on the spectral detuning from resonance and the atomic density.  At the densities where the measurements are performed, the dipole-dipole interaction plays an important role in the system physics, and leads to a significant variation in the transparency of the atomic sample with increasing density.  We emphasize that these studies are importantly different than previous investigations in that the coherently forward scattered light is measured, in contrast to the radiation diffusely scattered in other directions \cite{Pellegrino}.  For forward scattering the absolute light scattering cross section is also determined.

\begin{figure}
{\includegraphics{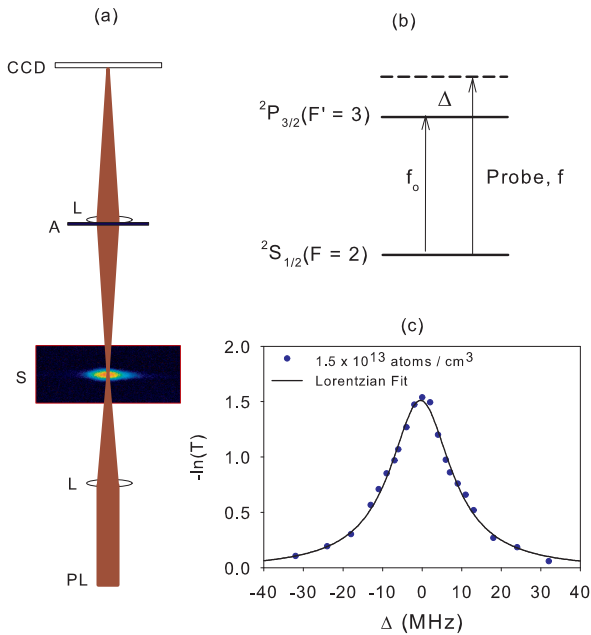}}
\caption{(a) Schematic layout of the experimental arrangement.  Here CCD refers to a charge coupled device detector, L a lens, A an aperture, PL the probe laser, and S the atom sample.  (b) A partial energy level of the pertinent transitions in $^{87}Rb$.  The hyperfine components are indicated in parenthesis.  $\Delta$ = $f - f_o$ is the detuning from atomic resonance at frequency $f_o$. (c) A representative spectral scan of transmission of the probe laser through the atomic sample. The ordinate displays the measured value of -lnT, where T is the spatially integrated transmitted probe intensity, normalized to the incident value. }
\label{Figure1}
\end{figure}

Samples of dense and cold $^{87}$Rb gas are prepared in a multistep process.   First, atoms from a warm atomic vapor are cooled and loaded into a magneto optical trap (MOT) \cite{Metcalf}, which operates near the 5s $^2S_{1/2}$ $F = 2  \to$   5p $^2P_{3/2}$ $F^{\prime} = 3$ hyperfine transition. The trapping laser is derived from an external-cavity diode laser (ECDL) system in a master-slave configuration, and is split into three pairs of retro reflected beams, delivering an estimated trapping laser intensity of $ \simeq 40$ mW/cm$^2$.  To prevent off-resonance optical pumping of the atomic vapor into the lower energy $F = 1$ hyperfine level, a repumping laser is set to the 5s $^2S_{1/2}$ $F = 1 \to$ 5p$^2P_{3/2}$ $F^{\prime} = 2$ hyperfine transition. The trap and repumping lasers are frequency tuned and rapidly switched on or off through acousto optical modulators (AOM). Fluorescence imaging measurements show that the sample has a nearly Gaussian spatial atom distribution of peak density $\rho_o$, described by $\rho(r) = \rho_o e^{-r^2/2r_o^2}$; for this sample $r_o \approx 0.4$ mm.  The temperature of the $^{87}$Rb atomic gas was determined via ballistic expansion to be $\sim 200\ \mu$K.  A maximum optical depth of $b \sim 10$ on the trapping transition was determined through absorption imaging measurements, resulting in a peak atomic density of $\rho_o \sim 9$ x $10^{10} $ atoms/cm$^3$.  The number of atoms in the MOT is determined from the peak optical depth and the sample dimensions as determined by fluorescence imaging.   This critical value is also found to agree within $10 \%$ with measurements employing a robust optical pumping method \cite{Chen}.  Second, some of the atoms in the MOT are loaded into a far off resonance optical dipole trap (FORT) operating at a power of 2.1 W and a wavelength of 1.064 $\mu m$.  The laser is focused to a beam spot size $\sim$ 18 $\mu m$ and has a Rayleigh length of 0.9 mm, resulting in a trap depth of about 700 $\mu K$.  The FORT is spatially overlapped with the MOT.  To accomplish efficient FORT loading, the cold atomic cloud is dynamically compressed and further cooled through detuning the MOT lasers by about -60 MHz and simultaneously reducing the repumper laser power by $\sim$ 100 times.  The MOT and repumper lasers are then extinguished and the quadrupole magnetic field is shut off. The entire process of loading the atoms into the F = 1 hyperfine level takes approximately 70 ms. A waiting period of $\sim$ 200 ms is required for the atoms in the FORT to come to thermal equilibrium at $\sim 100\ \mu$K and a peak density $\rho_0 \sim 2$ x $10^{13}$ atoms/cm$^3$. The atomic sample is well described by a bi-Gaussian spatial distribution of atoms, in the F = 1 ground hyperfine level, having Gaussian radii of $r_o$ = 3.6 $\mu m$ and $z_o$ =  280 $\mu m$.

In the experiments reported here (see Fig. 1(a)), we measure the forward scattered intensity of a near resonance probe laser beam having a wavelength $\sim$ 780 nm, and tuned in a range $\pm$ 30 MHz about the 5s $^2S_{1/2}$ $F = 2  \to$  5p $^2P_{3/2}$ $F^{\prime} = 3$ nearly-closed hyperfine transition.  The natural width of this transition is about 6.1 MHz. The probe beam bisects the biGaussian atom sample, and is focused near the center of the atom cloud.  The probe beam profile is well represented as a focused Gaussian beam with an intensity waist of 2.5 $\mu m$ and a Rayleigh range of 99 $\mu m$.  The laser power of the linearly polarized probe is about 75 pW, giving an on resonance saturation parameter of $s_o$ $\sim$ 0.1, meaning that saturation effects should be minimal in the experiments reported here.  Note that with the relatively tight focus, the probe beam, when tuned to atomic resonance, is strongly occluded by the atomic sample.   Upon exiting the sample chamber, the probe beam is passed through a variable circular aperture and then focused to a light detector using an achromatic doublet lens arranged in a unit magnification 2f - 2f configuration; here f is the focal length of the collection lens.  The detector used is a charge coupled device (CCD) camera, which collects the spatially distributed intensity optically transferred from the sample to the detector plane.

\begin{figure}
{\includegraphics{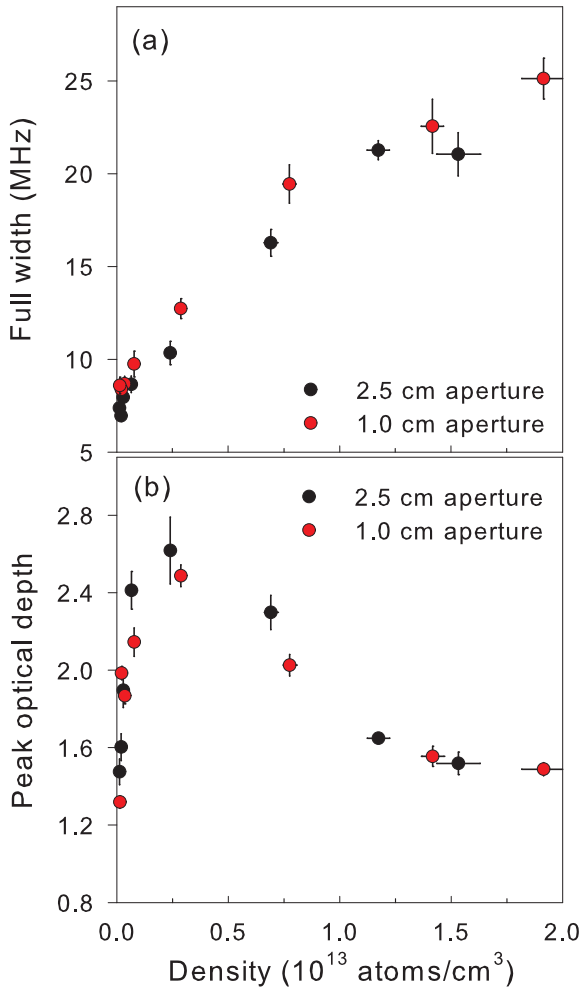}}
\caption{Density variations of observables for forward scattering.  (a) Full width at half maximum of the spectral dependence of the optical depth. (b) Peak optical depth, defined as -ln(T) where T is the normalized transmission.  }
\label{Figure2}
\end{figure}

\begin{figure}
{\includegraphics{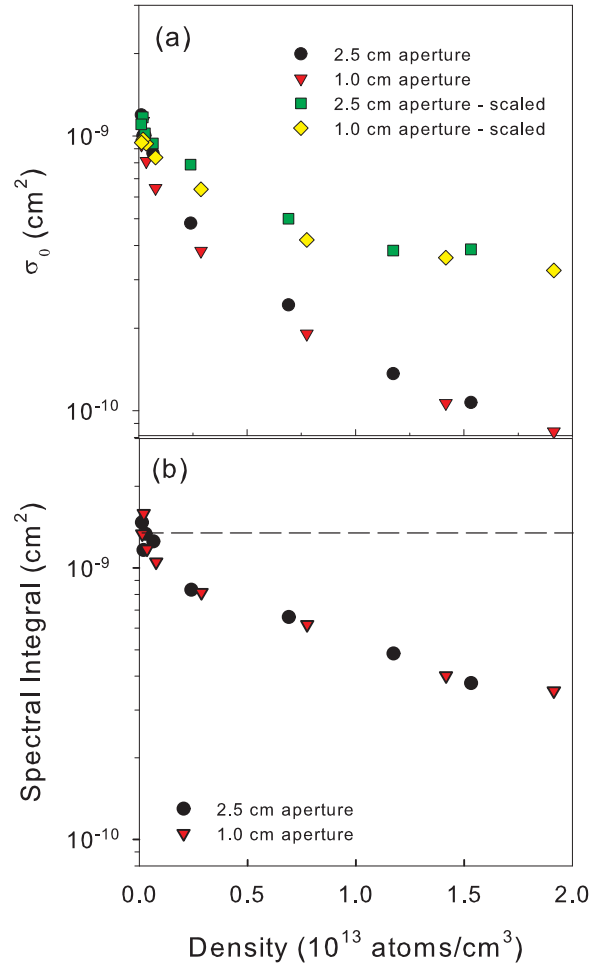}}
\caption{Density variations of the forward scattering cross section.  (a) Measured and rescaled forward scattering cross section. (b) Spectral integral.  On theoretical grounds, this integral is independent of density.  }
\label{Figure3}
\end{figure}

Once the sample is constituted, it is confined in the FORT until measurements are made.  Then the FORT laser is extinguished, a process which takes several $\mu s$.  After a variable delay time, during which the sample ballistically expands, the atoms are optically pumped to the F = 2 ground hyperfine level, a process which takes approximately 8 $\mu s$.  Then the sample is exposed to a probe laser beam for a period of about 10 $\mu s$. Ballistic expansion before probing controllably reduces the density at a nearly fixed total number of atoms.  Exposure of the sample to the probe beam after some delay then corresponds to a measurement at reduced density, and increased sample size.  As the decrease is quite slow on the scale of the measurement time, the density is nearly static on the measurement time scale.  In an alternate method, the atomic sample is held in the FORT for increasingly long delay times, during which the number of atoms decreases due to depopulation of the FORT by background gas collisions.  Then the FORT laser was turned off, and the transmitted light intensity measured by the integrated signals detected on the CCD camera. Each of these methods yielded similar results; in the present letter we focus on the ballistic expansion method.  In addition, we make measurements for different size light collection apertures (Fig. 1(a)).  This is done to assess the role of mode mismatching of the forward scattered light and the incident probe beam; the data presented here shows only a weak dependence on this effect.  Tests were made to determine any level of forward scattered light in the linear polarization channel $\emph{orthogonal}$ to the coherent one; if present, diffusely scattered light should be unpolarized and thus contribute to the signal.  Within the sensitivity of our measurements, no such signal was seen, insuring that incoherently forward scattered light is negligible here.

The main experimental results consist of measurements of the density and spectral dependence of light forward scattered from a $^{87}Rb$ atomic ensemble. Here we report traditionally defined values of the optical depth b = - lnT, where T is the measured light intensity normalized to the incident intensity in the absence of the atomic ensemble. For transmission of light through the center of a Gaussian atomic cloud, the optical depth $b$ = $\sqrt{2\pi} \sigma \rho_{o} r_{o} $, where $\rho_{o}$ is the atomic density in the center of the cloud, and $r_{o}$ is the Gaussian radius of the cloud. $\sigma$ is the cross section, assuming that it does not depend on atomic density.  A characteristic spectrum obtained at a peak density of $1.5$ x $10^{13}$ $atoms/cm^{3}$ is shown in Fig. 1(c).  The data are very well fit by a Lorentzian, from which the frequency offset from bare atomic resonance, the full width at half maximum (FWHM) of the response, and the peak optical depth is determined.  First we consider the the measured frequency offset; for this quantity, we are able to set only an upper technical limit of about $\pm$ 1 MHz.  Second, we present in Fig. 2(a) our results for the width of the spectral response. Here we see a significant increase of the FWHM with increasing density; at the highest density the FWHM is about 25 MHz.  At the lowest attained densities, the width is slightly larger than 6.1 MHz, the natural width of the transition. This difference is likely due to a combination of the probe spectral width, the residual Doppler width of the atomic resonance, and the long term stability of the central laser frequency.  We point out that the increasing width for our spatially larger samples should primarily be a density, rather than optical depth dependent, effect; we have confirmed this by separate experiments on samples of increasing density, but fixed sample size, where the FWHM is consistent with the results of Fig. 2(a).  As for the overall trend of the data, we see an increase in the FWHM, but a tendency towards a slowing down of the increase with density.  In comparing these results with earlier measurements \cite{Pellegrino}, we see that the overall magnitude of the FWHM is similar, and there is a similar tendency towards saturation.  Absolute comparison cannot be directly made because the measurements of \cite{Pellegrino} (a) are made on a microscopic sample for which a larger number of atoms are in the proximity of the surface, and (b) the spectral response of the diffusive component of scattered light is measured, rather than the forward scattered light.

We now turn to the density dependence of the peak optical depth, as shown in Fig. 2(b).  The overall nonmonotonic behavior observed is due to several factors that appear in the expression for the optical depth b. First, the atomic density is varied by ballistic increase of the value of the Gaussian radius, $r_{o}$.  This implicit density dependence, and the overall density dependence itself, suggests expressing the basic measurements of the optical depth in terms of a light scattering cross section $\sigma$.  This we do in the lower set of data points (circles and inverted triangles) in Fig. 3(a).  We point out that the cross section shown there is an effective one, averaged over the overlap between the inhomogeneous atomic cloud and probe beam.  In the figure, we see that the cross section regains monotonic behavior and, for the lowest densities, approaches the expected isolated atomic cross section of $1.36$ x $10^{-9}$ $cm^{2}$ for this transition.  However, we also see that the resulting cross section decreases by $\sim$ an order of magnitude as the density is increased to the largest attained value. This sharp decrease is due to several factors. The primary experimental cause is the fact that the laser probe profile as it crosses the atomic cloud samples an average density smaller than the maximum $\rho_o$ at the cloud center, resulting in increased transmission.  Such an effect depends on the atomic density and on the atomic sample size.  Simulations of this averaging show, at the highest density, a factor of about three decrease in the measured transmission.  A second but less important effect is the absence of ideal mode matching between the incident probe field and the response field generated by the atoms in the cold ensemble.

It is useful to have a second and independent determination of this effect.  Such a measure is illustrated in Fig. 3(b), where a spectral integral is graphed versus atomic density.  The normalized spectral integral, as we define it, is the cross section integrated over detuning (see Fig. 1(c)) and then divided by the natural width.  Crucially, recent rigorous calculations \cite{Sokolov1,Sokolov2} have shown that this integral, for the case of forward scattered light,  is constant over a wide range of densities, including that explored here.  This constraint allows us to use the results of Fig. 3(b) to rescale the density dependent cross section.  The result of this rescaling is shown in the upper points (diamonds and squares) of Fig. 3(a).  The clear remaining decrease is the main result of this paper, which is the suppression, and the absolute magnitude of, the cross section for forward scattering as a function of atomic density.  The overall size of this effect is in qualitative agreement with recently reported fluorescence measurements \cite{Pellegrino}, but are put on an absolute scale.

Recent research \cite{Sokolov1,Sokolov2,Kaiser,Balik,Pellegrino} gives a framework for understanding the main result of this paper.  First all, from a physical point of view it is important to realize that the average atom-atom separation is small enough that the shorter range dipole-dipole interactions are also important  \cite{Bouloufa}, but that the long-range radiative part of the dipole dipole interaction also drives larger length scale cooperative interactions among the system particles \cite{Kaiser}.  The deviation from linear density dependence of the optical depth at higher density is then caused by interactions among the system particles via the interatomic resonant dipole–dipole interaction. This interaction generally leads to coupling among the atoms in the ensemble, and particularly generates shifts of collective resonances, so that the peak cross section for resonance radiation decreases. Then the overall increase in the peak optical depth due to the linear density dependence in the Beer Lambert Law is partially suppressed by the resonance shifts and broadening. Overall, a quasimodal analysis of the width and location of the poles of the resolvent \cite{Balik,Sokolov1,Sokolov2} is also useful. These modes may be considered as sub- or super-radiant depending on whether the associated width is smaller or greater than the 6.1 MHz natural decay width of the free atom.  In our experiment, the spectral profile of the probe matches better with larger width modes, leading to preferential excitation of superradiant-type modes, and an associated spectral response of increased width, as observed in experiment \cite{Balik,Pellegrino}.

In conclusion, we have detailed experimental investigation of collective density effects on coherent forward scattering of light in cold atomic gases. The measurements show that there is a substantial suppression of the peak light scattering cross section with increasing density, leading to an increased transparency of the atomic cloud.  Interpretation of the measurements in terms of the constancy of a spectral integral with density permits the cross section measurements to be put on an absolute scale.   The results may be understood by considering the optical excitation to be described in terms of quasimodes of the system, whose cooperative behavior lead to both broadening and suppression of the peak cross section.

We appreciate financial support by the National Science Foundation (Grant Nos. NSF-PHY-0654226 and NSF-PHY-1068159), the Russian Foundation for Basic Research (Grant No. RFBR-CNRS 12-02-91056). D.V.K. would like to acknowledge support from the External Fellowship Program of the Russian Quantum Center (Ref. Number 86). We also acknowledge the generous support of the Ministry of Education and Science of the Russian Federation (State Assignment 3.1446.2014K).

\end{document}